\documentclass{article}
\usepackage{graphicx}
\usepackage{subcaption}
\usepackage{geometry}
\usepackage{float}
\usepackage{xcolor}
\usepackage{colortbl}
\usepackage[T1]{fontenc}
\usepackage{booktabs}
\usepackage{siunitx}
\usepackage[
    backend=biber,
    style=apa,
    natbib=true,        
    sorting=nyt         
]{biblatex}
\usepackage{hyperref}
\hypersetup{colorlinks=true, urlcolor=blue} 

\usepackage{setspace}

\title{Robust Bioacoustic Detection via Richly Labelled Synthetic Soundscape Augmentation}
\author{
    Kaspar Soltero\textsuperscript{1*}, Tadeu Siqueira\textsuperscript{2}, Stefanie Gutschmidt\textsuperscript{1} \\ \\ 
    \textsuperscript{1}Department of Mechanical Engineering, University of Canterbury, \\ Christchurch, New Zealand \\
    \textsuperscript{2}Institute of Biosciences, Sao Paulo State University (UNESP), \\ Rio Claro, Brazil\\ \\
    \textsuperscript{*}Corresponding author: \href{mailto:kaspar.soltero@pg.canterbury.ac.nz}{kaspar.soltero@pg.canterbury.ac.nz} 
}
\date{July 2025} 

\begin{document}
\maketitle

\begin{abstract}
1. Passive Acoustic Monitoring (PAM) generates large datasets valuable for ecological assessment, but analysing these data, particularly for vocalisation-level detection, is often hindered by the intensive manual effort required to create richly labelled training datasets. This study introduces and evaluates a novel synthetic data framework designed to generate large volumes of richly labelled training data from very limited source material, aiming to improve the robustness and generalisability of bioacoustic detection models.\\

2. We developed a sample-augment-combine framework that synthesises realistic 10-second soundscapes by combining segments of clean background noise with carefully isolated target vocalisations (little owl, \textit{Athene noctua}) and contaminant sounds. Crucially, bounding box and segmentation mask labels for target vocalisations are generated automatically and dynamically during synthesis, accounting for random placement, signal-to-noise ratio adjustments, and background augmentations. We used this synthetic data to fine-tune an image classifier model (EfficientNetB0 pre-trained on the common ImageNet).\\

3. By evaluating model performance on a held-out dataset of real-world soundscapes using the soundscape-level Area Under the Curve (AUC) metric and F1 score, we found that model performance generalised well to the real-world domain. Furthermore, the evaluation metrics remained high even when the diversity of unique source vocalisations used for synthesis was drastically reduced, suggesting that the model learned generalised features rather than overfitting to augmentation artifacts.\\

4. Our results demonstrate that richly labelled synthetic data generation is a highly effective strategy for training robust bioacoustic detectors from small source datasets. This approach significantly reduces manual labelling effort, overcomes a key bottleneck in computational bioacoustics, and facilitates the development of high-resolution acoustic monitoring tools potentially applicable to diverse species and environments, thereby enhancing ecological assessment capabilities.\\
\end{abstract}

\newpage
\section{Introduction}
In recent years, 'Passive Acoustic Monitoring' (PAM) microphone networks have made possible new forms of ecological assessment, offering a less invasive and more scalable approach compared to traditional, manual methods \citep{lostanlenRobustSoundEvent2019}. Improving microphones, reducing costs, and the ability to permanently store and re-analyse data has made PAM increasingly preferred by researchers over manual sampling \citep{stevensonGeneralFrameworkAnimal2015}. Acoustic monitoring (as compared with other passive monitoring methods like satellite imagery or camera traps) is suited to this purpose because it can provide high-resolution data across broad spatiotemporal scales \citep{shamonUsingEcoacousticsMetrices2021}, with an increase in data volume generally resulting in a more accurate assessment \citep{sugaiDriversAssemblagewideCalling2021, wuEvaluatingCommunitywideTemporal2024}. One of the most promising applications of PAM is in terrestrial community ecology, through animal population monitoring \citep{sugaiTerrestrialPassiveAcoustic2019a}. In response to increasingly large volumes of PAM data, a growing body of work seeks to address the problem of automatically analysing soundscapes to query ecological information of interest \citep{kvsnBioacousticsDataAnalysis2020}. There remains a critical bottleneck in the analysis of PAM data, however, due to the intensive manual effort required to create richly labelled training datasets.
\\

The most advanced PAM analyses utilise machine learning algorithms to automatically label sounds of interest within soundscapes. Machine learning approaches generally separate this analysis into two tasks, which are often applied in series: detection and classification \citep{stowellComputationalBioacousticsDeep2022}. In this context, detection refers to detecting a specific sound event within a broader soundscape (essentially a binary task), and classification refers to discerning a 'class label' for a given sound event. Both tasks may be conducted to varying degrees of specificity. For example, a low-specificity classification model may discern broadly whether a soundscape contains anthrophony (e.g. vehicles, drones), biophony (e.g. animal vocalisations), or geophony (e.g. rain, wind) \citep{quinnSoundscapeClassificationConvolutional2022}, while the highest-specificity 'censusing' models may be used to distinguish individual animals of the same species \citep{guerreroAcousticAnimalIdentification2023}. For the detection task, specificity refers to how accurately the signal is pinpointed within the soundscape. Detection methods range from low-specificity such as presence-absence within the soundscape \citep{lostanlenRobustSoundEvent2019} or time-domain only
\citep{hoffmanSyntheticDataEnables2025,venkateshYouOnlyHear2022} detection, to high-specificity such as time-frequency box object detection \citep{fanioudakisDeepNetworksTag2017, zsebokAutomaticBirdSong2019, wuSILICCrossDatabase2022} or time-frequency mask detection \citep{izadiSeparationOverlappingSources2020}.
\\

High-specificity vocalisation-object detection algorithms are a recent architectural development in bioacoustic machine learning. Compared with low-specificity detection architectures, higher-specificity vocalisation detection offers several advantages. Whereas soundscape-level detection is sufficient for estimating presence-absence based metrics such as diversity indexes, the higher temporal specificity of box object or mask detection allows for more ecologically useful metrics like the Vocalisation Activity Rate (VAR) to be directly inferred \citep{hutschenreiterHowCountBird2024}. The additional frequency range detection not only provides information for simple frequency filtering, but may also improve downstream classification performance for small objects (vocalisations with small frequency or time spans) by focusing the classification task on the relevant frequency bands / time windows \citep{wuSILICCrossDatabase2022, jiangReviewYoloAlgorithm2022}. Unfortunately manually creating richly labelled training data is a potential source of bias. More critically, the manual data-labelling process is often prohibitively intensive for training box and mask detection networks \citep{xieReviewAutomaticRecognition2023}.
\\

Even lower specificity bioacoustic machine learning methods face the challenge of data scarcity. Many machine learning techniques which have found success in data-rich fields like speech processing encounter a lack of labelled training data when applied to bioacoustics \citep{mcewenAutomaticNoiseReduction2023}. While advanced classification models such as PERCH and BirdNet can classify hundreds of animal species with a relatively high accuracy \citep{ghaniGlobalBirdsongEmbeddings2023}, often animals of interest fall outside this narrow subset of species. Bioacoustic machine learning models are typically designed for specific contexts, and the broad acoustic diversity of PAM surveys inhibits their transferability and generalisation \citep{stowellComputationalBioacousticsDeep2022}. This diversity may include novel ecosystems and animal species, novel background noise characteristics, or novel recording equipment and preprocessing methods compared with the training environment of the model \citep{merchanBioacousticClassificationAntillean2020}. Tasks demanding higher-specificity detections (such as extraction of VAR) necessitate both higher intensity data labelling and often larger training datasets. Therefore, developing well-generalised models for these tasks is inherently more challenging compared to soundscape-level networks.
\\

The two key techniques for overcoming small bioacoustic training dataset size are data augmentation and transfer learning
\citep{gulSurveyAudioEnhancement2023}. Transfer learning refers to the training of networks which have first been pretrained on data from an adjacent domain \citep{baptistaBioacusticClassificationFramework2021}. To effectively make use of the pretrained features, the new dataset must be preprocessed into a representation which considers the differences between the two domains \citep{gulSurveyAudioEnhancement2023}. Image processing is the preferred domain on which to pretrain bioacoustic models, because audio can be represented in the spectral domain (e.g. via the Short-Time Fourier Transform (STFT)) and easily transformed into an image-like spectrogram format \citep{zhongMultispeciesBioacousticClassification2020}. Careful consideration must be given to maximise the transfer of features from the image domain to the audio domain, such as ensuring the same normalisation regime \citep{ghaniGlobalBirdsongEmbeddings2023}.
\\

Data augmentation is a widely used technique to apply artificial variation and increase the size of training datasets 
\citep{kahlBirdNETDeepLearning2021}. Commonly applied augmentations include the addition of random noise, shifts in time, the 'mixing' together of training examples, and random masking or deletion of information \citep{herbstEmpiricalEvaluationVariational2024}. The success of augmentation methods is domain-dependent 
\citep{stowellAutomaticAcousticIdentification2018}. For example, vertical flipping and rotation are common augmentations in the image domain \citep{hatayaMetaApproachData2022}, but may not be well suited to audio-spectrogram tasks because inverting the frequency or time axis does not produce a physically realistic soundscape. Excessive data augmentation can hinder model performance, possibly by the introduction of augmentation artifacts, leading downstream models to overfit to these artifacts rather than generalise to unseen data \citep{macisaacImprovingAcousticSpecies2024}. The implied goal of augmenting a dataset to a larger size is to increase the robustness of the model's learned features to the highly variable real-world testing environment. Therefore, bioacoustic data augmentation should aim to produce novel-seeming and realistically labelled soundscape data \citep{sunClassificationAnimalSounds2022}. 
\\

Data augmentation can extend beyond the simple transformation of existing samples to include the generation of entirely new samples or 'synthetic' data. Synthetic data techniques often rely on generative models such as Generative Adversarial Networks (GANs) or diffusion-based models to produce new training examples \citep{fuClassificationBirdsongSpectrograms2023}. \citet{herbstEmpiricalEvaluationVariational2024a} show that generative synthetic methods offer only mild performance improvements while requiring significantly higher computational cost. \citet{gibbonsGenerativeAIbasedData2024} note that these generative methods can produce distorted data, and do not offer controllable generation. An alternative to generative models is the sample-augment-combine approach proposed by \citet{hoffmanSyntheticDataEnables2025}. This approach involves preprocessing and augmenting existing samples before combining them to create new training examples. The sample-augment-combine method is less computationally expensive than generative models, however it still requires careful consideration of the augmentation methods used to ensure that the generated data is representative of real-world conditions.
\\

Despite advances in bioacoustic machine learning, critical challenges persist in developing high-resolution vocalisation detection models that generalize across diverse acoustic environments. To address some of these gaps, we present a novel synthetic data augmentation framework that dynamically generates large volumes of realistically augmented soundscapes with automatically computed bounding boxes and segmentation masks, eliminating manual labelling effort while preserving acoustic plausibility. By systematically combining isolated vocalisations with variable background noise and contaminant sounds at controlled signal-to-noise ratios, our approach enables the creation of training datasets that capture the spectral and temporal complexity of real-world soundscapes. We hypothesize that training with synthetic data incorporating lower SNRs will improve model robustness to low-SNR vocalisations in real-world data. We also expect, based on \citet{macisaacImprovingAcousticSpecies2024}, that increasing the synthetic dataset size will improve model performance up to a point, after which the model will begin to overfit to the introduced artifacts and generalisation performance will decrease. Through rigorous evaluation of synthetic generation and source diversity effects, we provide insights into balancing synthetic dataset scale and ecological validity. Our approach both reduces dependency on labour-intensive manual annotations and provides a flexible framework for developing species-adaptive detection tools, advancing the potential of PAM to deliver high-resolution ecological insights across taxonomic groups and ecosystem types.

\section{Materials and Methods}
\subsection{Datasets}
Four datasets were used in the study; three for training and one for evaluation. The first three datasets (A, B, and C) contained 'clean background noise', 'isolated sounds of interest', and 'isolated contaminant sounds' respectively. These were sourced both privately and from the online Xeno-Canto database \citep{XenoCanto2025}, attributions for which may be found in the associated repository. Datasets A, B, and C were sampled, augmented, and mixed together to form the training data. Dataset B contained only vocalisations with a high Signal-to-Noise-Ratio (SNR), which allowed the training data to be generated with a wide variety of SNRs as described in Section \ref{sec:data_augmentation}. The fourth dataset (D) was used for evaluation, consisting of low-resolution, manually labelled soundscape data.
\\

The background noise dataset (A) was collected both privately and from Xeno-Canto soundscape recordings. Segments were selected for containing no audible animal vocalisations for a duration longer than 10 seconds. Background noise segments were checked carefully for existing animal vocalisations, to ensure that transformations during the later augmentation process would not draw out faint vocalisations which were initially inaudible. Transient contaminant noise, such as categories included in dataset C, were allowed also in the 'clean' background noise dataset. This is because contaminant noises represent realistic background conditions which downstream models should be robust to. Maximum and minimum frequency bounds were associated with each background segment, based on their different recording frequencies and potential prior band-pass filtering.
\\

\begin{figure}[t!]
    \centering
    \includegraphics[width=\linewidth]{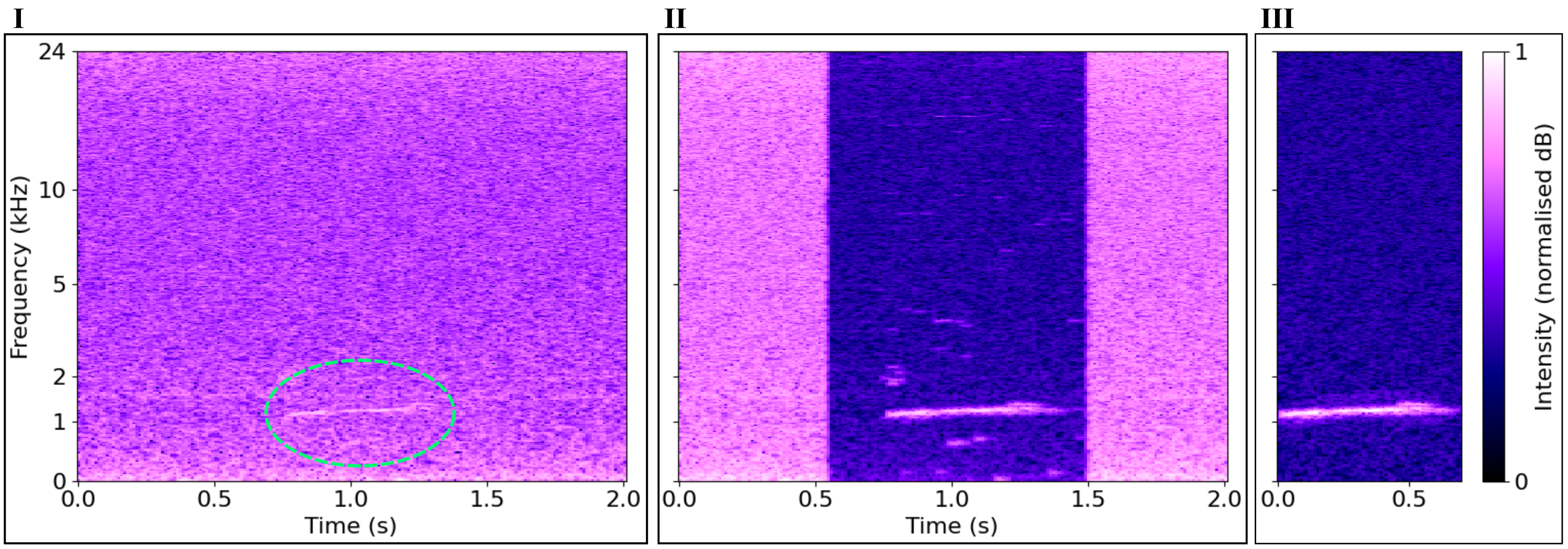}
    \caption{Spectral extraction of a vocalisation from the background noise. (\textbf{I}) The original soundscape segment with the vocalisation highlighted. (\textbf{II}) The original soundscape after spectral subtraction. (\textbf{III}) The isolated vocalisation after mask-cropping.}
    \label{fig:spectral_subtraction}
\end{figure}

The isolated sounds of interest dataset (B) consisted of recordings of \textit{Athene noctua} (Little Owl) collected from Canterbury, New Zealand. The preparation of this dataset is the most critical piece of the methodology. This is a required part of the framework and is the most manual component. To extend our method to other species, only this dataset would require additional data. Vocalisations were selected for 1) being clearly distinguishable by ear in the recording without overlapping contaminant sounds, 2) for having a period relatively free of any vocalisations nearby them in the recording, and 3) for having stationary background noise (without transient noise such as wind gusts). This nearby period, free of animal vocalisations, was used as an estimate of the stationary background noise's spectral profile to perform spectral subtraction over the vocalisation. Following spectral subtraction, the vocalisation was carefully cropped from any remaining background noise in the spectral domain in Adobe Audition, to produce an approximation of the original signal (Figure \ref{fig:spectral_subtraction}). Finally, isolated vocalisations below an RMS of 70 dBFS were discarded to ensure a clear enough signal for augmentation. The final curated dataset consisted of 30 unique vocalisations of a single sonotype.
\\

The isolated contaminant sounds dataset (C) was formed in the same manner as the isolated vocalisation dataset, with the exception that it consisted only of sounds from non-interest species, as well as transient geophonic and anthrophonic sources such as wind, rain, cars, and planes. This dataset contained 58 isolated sounds encompassing a range of durations and frequencies. The evaluation dataset (D) consisted of 10-second soundscape recordings manually labelled as either positive or negative for vocalisations of \textit{Athene noctua}. These recordings were collected from a PAM survey in Canterbury, New Zealand from August 2024 - January 2025 and labelled by a human expert. The evaluation dataset contained 74 soundscape segments, with 24 positive and 48 negative examples.\\

\subsection{Data Augmentation}\label{sec:data_augmentation}
\begin{figure}[htbp]
    \centering
    \includegraphics[width=\linewidth]{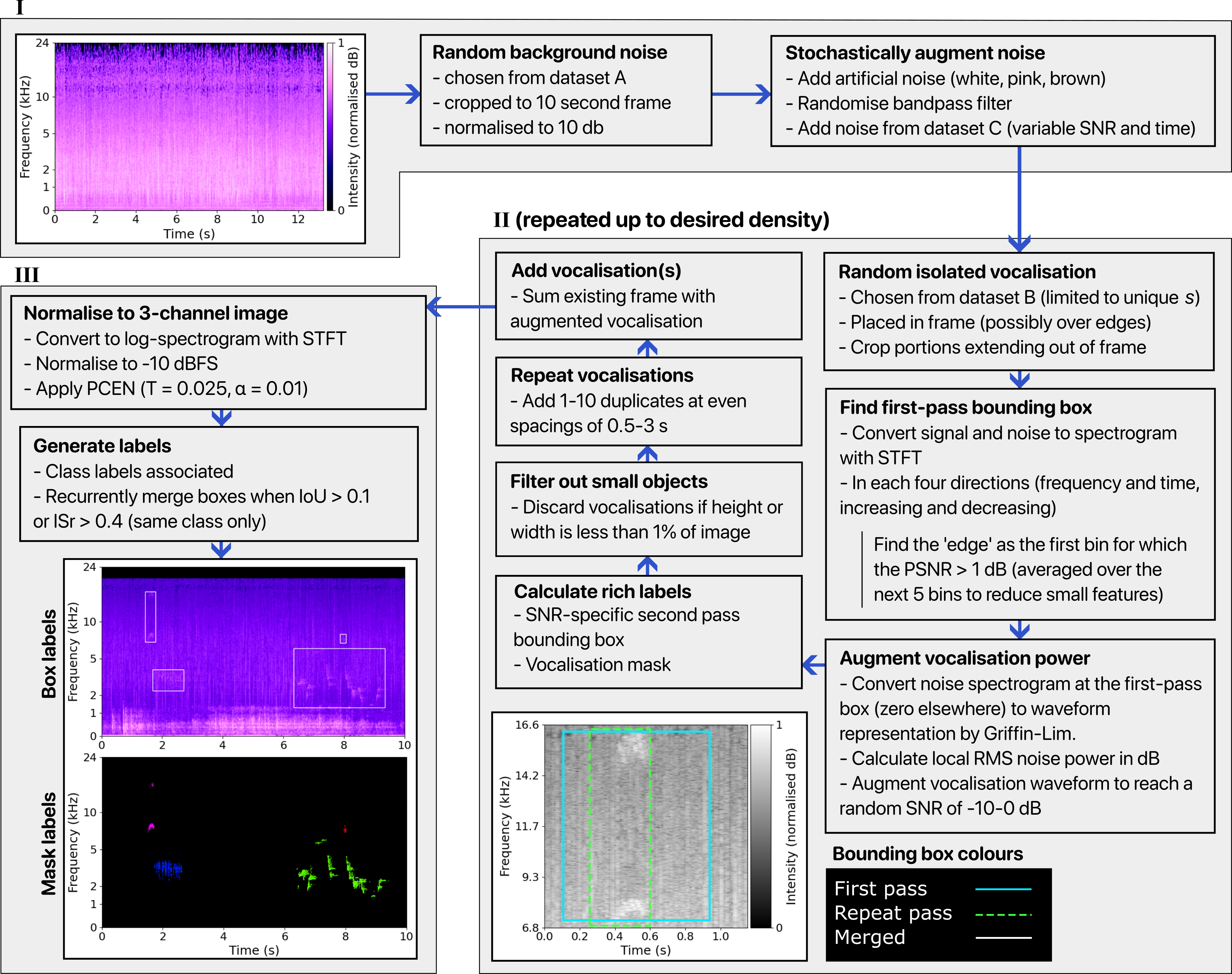}
    \caption{Data augmentation process for creating each 10 second artificial soundscape. \textbf{(I)} A clip of real background noise is stochastically augmented and overlaid with contaminant sounds. \textbf{(II)} Isolated vocalisations are added to the background at varying SNR, with dynamically calculated bounding boxes and vocalisation masks, until desired vocalisation density is reached (we range density from 0-2). \textbf{(III)} Final artificial soundscape is normalised and class labels at the soundscape, box, and mask level are stored (vocalisation mask colours represent distinct classes).}
    \label{fig:methods_diagram}
\end{figure}

The training dataset was formed by augmenting and mixing datasets A, B, and C to produce a wide variety of realistic, synthetic soundscapes. While the training dataset represented a single-class task, Figure \ref{fig:methods_diagram} shows the data augmentation process using multiple classes to demonstrate the rich-labelling extensibility of the method.\\

Random background noises and \textit{Athene noctua} vocalisations were mixed together at a vocalisation density between 0 and 2, to automatically generate richly labelled 10-second training examples. Segments were randomly cropped from dataset A and added with random amounts of gaussian noise, and contaminating segments from dataset C, to simulate diverse background noise. \textit{Athene noctua} vocalisations were mixed into this background noise at a random SNR, while simultaneously their rich bounding labels were calculated. By using approximations of the original vocalisation signals (dataset B), the spectral bounding box and precise mask of each vocalisation in the artificial soundscape could be dynamically determined to a high accuracy during dataset creation. In other words, a single vocalisation is re-used many times and the bounding box and mask are calculated each time based on its placement, its power adjustment, and the augmentations applied to the background noise, rather than being fixed. The STFT is used in two parts of the augmentation process, when calculating the vocalisation SNR adjustment and when creating the final image. In both cases, the STFT is used to create a power-spectrogram with FFT value 2048, window size 2048, and hop length 512.\\

Generating the labelled data automatically removes the need for tedious manual labelling and allows thousands of training examples to be generated from relatively little input. This has the added benefit of producing consistent labels, preventing observer bias which could potentially interfere if bounding boxes or masks were recorded manually. In case of vocalisations obscuring each other, bounding boxes of the same class are merged if their ratio of intersecting area to union area (IoU) is greater than 0.25, or intersecting area to the area of the smaller box (IRs) is greater than 0.9 as in \citet{wuSILICCrossDatabase2022}. Artificial soundscapes were generated as 10-second segments, to balance the need for soundscapes to be large enough to encompass a variety of background noise and vocalisations, while still being small enough to see small vocalisations in the spectrogram. We generated synthetic datasets of varying sizes ($n$), as well as datasets with varying total samples of isolated vocalisations from dataset B ($s$), to evaluate the effect of augmentation diversity on model performance. We also generated datasets with varying SNR ranges, to evaluate the controllability of the synthetic dataset and its effect on the model's robustness to low SNR vocalisations.

\subsection{Preprocessing}
Prior to ingestion by the image model, both artificial and real-world (RW/evaluation) soundscape segments were pre-processed in the same manner. All audio was resampled to 48 kHz, normalised to -10 dB, then transformed to a power spectrogram using the STFT (Figure \ref{fig:methods_diagram}, \textbf{III}). STFT transformation used a window size of 2048, hop length of 512, and FFT size of 2048. Spectrograms were transformed into the appropriate image format by per-channel-energy-normalisation according to \citet{lostanlenPerChannelEnergyNormalization2019}, by shifting the frequency axis to log scale to better separate the lower frequency bands, and finally by Lanczos resampling to the necessary image dimensions of 256x256 pixels.

\subsection{Transfer Learning}
Transfer learning was employed on the popular image classification network EfficientNet \citep{tanEfficientNetRethinkingModel2020}. We used the smallest EfficientNet model, EfficientNetB0, pretrained on the ImageNet dataset. The model was fine-tuned on the synthetic training dataset at a resolution of 256x256 pixels. We fine-tuned the model for 20 epochs with a batch size of 32 using the Adam optimiser. We stopped the model training when the Area Under the Curve (AUC) on the real-world evaluation dataset did not improve for 8 epochs.

\subsection{Evaluation}\label{sec:evaluation}
Model evaluation followed the recommendations of \citet{knightRecommendationsAcousticRecognizer2017}. No audio present in dataset D were included in datasets A, B, or C. The evaluation dataset represented novel conditions and background noise characteristics compared to the training data. Although the data augmentation method is designed to be able to output high-specificity training data, suitable real-world evaluation datasets at this resolution are hard to come by. Therefore, we evaluated model performance using a soundscape-level binary classification task, where a soundscape was considered 'true positive' if it contained any vocalisation of \textit{Athene noctua}. We report the AUC and F1 metric at a threshold of 0.5.\\

We conducted three primary analyses of synthetic dataset performance. Firstly, we evaluated the effect of increasing the synthetic dataset size $n$ on model performance. We trained models with synthetic datasets ranging in size from 100 to 8000, and evaluated their performance on the held-out evaluation dataset. Secondly, we evaluated the effect of reducing the proportion of isolated vocalisations $s$ in the synthetic dataset on model performance. We trained models with synthetic datasets composed of the original 30 isolated vocalisations, ranging down to 1 isolated vocalisations, and evaluated their performance on the held-out evaluation dataset. For both parameter sweeps, we repeated the experiment four times to remove some of the stochastic effects introduced by the synthetic data generation framework.
\\

Our third analyis consisted of a qualitative evaluation of the effect of varying the SNR range of the synthetic dataset on model performance on low-SNR samples. We varied the SNR range of the synthetic dataset from 0.1-1 to 0.9-1 (linear scale) and evaluated the model performance on low-SNR samples from the evaluation dataset, in order to test the hypothesis that lower SNR minimas would improve model robustness to low-SNR vocalisations.
\\

\section{Results}
We investigated whether the data augmentation framework was beneficial to model performance. Increasing the augmented dataset size was beneficial to model performance up to $n = 1000$ (33x the number of unique samples) which reached an average AUC of 0.92 on the held-out evaluation dataset (see Table \ref{tab:parameter_sweep}a). From $n =$ 1000-8000, performance appeared to gradually decrease. This aligns with our hypothesis that the model would begin to overfit to the synthetic dataset as its size increased. However, the model performance remained high even with the largest synthetic dataset (266x larger than the original dataset), suggesting that although the model eventually began overfitting to introduced artifacts or biases, the augmentation framework was still effective at enhancing the model's generalization capability.

\begin{table}[htbp]
    \caption{Performance metrics (AUC and F1 score at a classification threshold of 0.5) for models trained on synthetic datasets with varying augmentation factors (left, $s$=30), and varying numbers of unique isolated vocalisations (right, $n$=1000). The mean and standard deviation from four training runs synthesised with the same settings are shown.}
    \centering
    \begin{subtable}{0.48\textwidth}
        \centering
        \caption{Varying augmentation factors ($s$=30)}
        \begin{tabular}{|c|c|c|}
        \hline
        $n$ & AUC & F1@0.5 \\
        \hline
        100 & $0.7461 \pm 0.0064$ & $0.5183 \pm 0.0068$ \\
        300 & $0.8362 \pm 0.0180$ & $0.6472 \pm 0.0312$ \\
        500 & $0.9060 \pm 0.0250$ & $0.7541 \pm 0.0587$ \\
        1000 & $0.9195 \pm 0.0140$ & $0.7758 \pm 0.0342$ \\
        2000 & $0.8924 \pm 0.0239$ & $0.7094 \pm 0.0421$ \\
        4000 & $0.8893 \pm 0.0124$ & $0.7216 \pm 0.0243$ \\
        8000 & $0.8865 \pm 0.0325$ & $0.7402 \pm 0.0338$ \\
        \hline
        \end{tabular}
        \label{tab:aug_factors}
    \end{subtable}
    \hspace{-10pt}
    \begin{subtable}{0.48\textwidth}
        \centering
        \caption{Varying unique vocalisations ($n$=1000)}
        \begin{tabular}{|c|c|c|}
        \hline
        $s$ & AUC & F1@0.5 \\
        \hline
        1 & $0.8535 \pm 0.0537$ & $0.5658 \pm 0.1030$ \\
        2 & $0.8722 \pm 0.0275$ & $0.6884 \pm 0.0666$ \\
        3 & $0.9054 \pm 0.0198$ & $0.7072 \pm 0.0433$ \\
        5 & $0.8902 \pm 0.0244$ & $0.6958 \pm 0.0666$ \\
        10 & $0.9043 \pm 0.0060$ & $0.7531 \pm 0.0199$ \\
        20 & $0.9115 \pm 0.0454$ & $0.7778 \pm 0.0647$ \\
        30 & $0.8911 \pm 0.0172$ & $0.6989 \pm 0.0314$ \\
        \hline
        \end{tabular}
        \label{tab:unique_vocal}
    \end{subtable}
    \label{tab:parameter_sweep}
\end{table}

We also investigated whether the model could learn and overfit to the underlying dataset of positive samples if this dataset size was small, by reducing the number of unique vocalisations from 30 to 1 while keeping the total synthetic dataset size constant at 1000 examples. Performance appeared to decrease gradually as the number of unique vocalisations was reduced, however even with only 1 unique vocalisation (0.1\% of the synthetic dataset size), the model still achieved an AUC of 0.85 on the held-out evaluation dataset (see Table \ref{tab:parameter_sweep}b). This suggests that the model was able to effectively learn generalised features from the synthetic dataset, rather than overfitting to the specific source data used. We also note that the model performance was more variable when only one unique vocalisation was used. This may indicate that vocalisation choice has a bigger impact on model performance at lower sample sizes.\\

\begin{figure}[htbp]
    \centering
    \includegraphics[width=\linewidth]{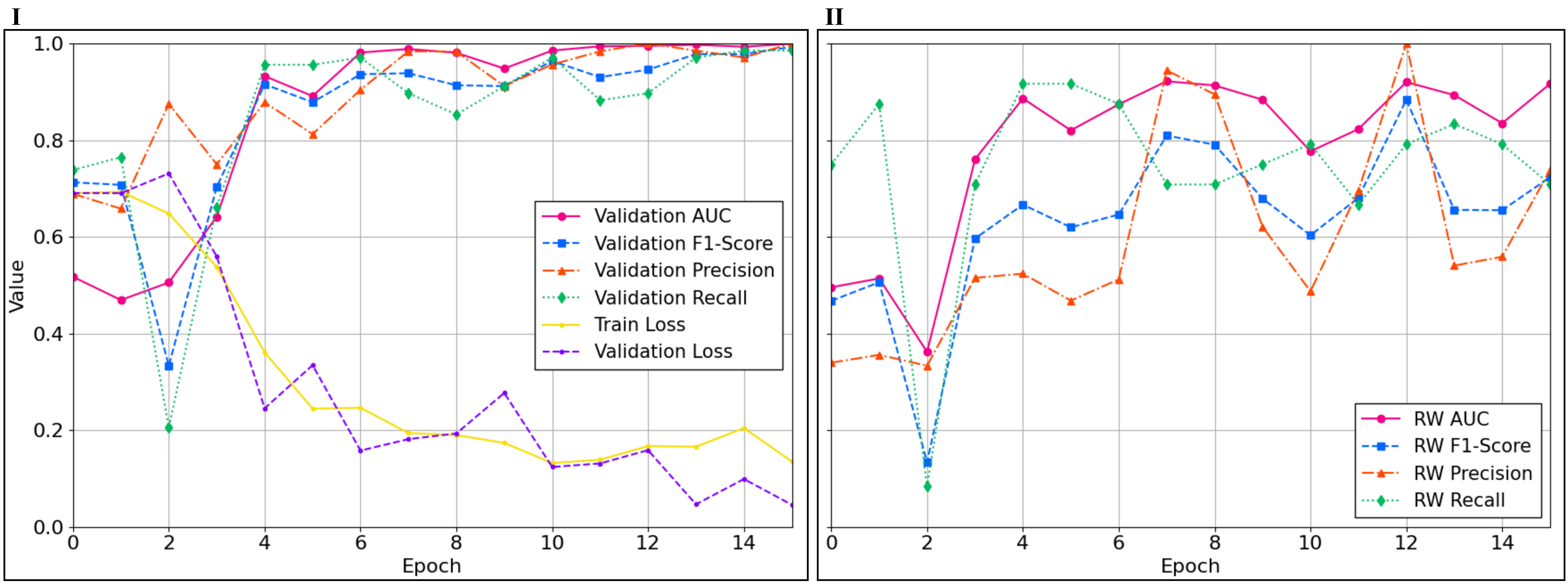}
    \caption{Training curves (best of 4 attempts) of EfficientNetB0 trained on a synthetic dataset with $n$=1000, $s$=30. (\textbf{I}) Loss curves and performance on the synthetic dataset. (\textbf{II}) Performance on the real-world (RW) evaluation dataset.}
    \label{fig:rl_metrics}
\end{figure}
We evaluated the AUC, F1, precision, and recall metrics during training of a model trained on the identified optimal synthetic dataset size of 1000 examples, with 30 unique vocalisations. The performance on the synthetic dataset was compared to the performance on the real-world evaluation dataset during training (Figure \ref{fig:rl_metrics}). Loss curves and performance during training showed the model successfully learns the synthetic dataset (Figure \ref{fig:rl_metrics}, \textbf{I}). Maximum performance on the real-world evaluation dataset show that the model generalises well to unseen data, with AUC and F1 scores of 0.92 and 0.88 respectively (Figure \ref{fig:rl_metrics}, \textbf{II}). There was, however, a clear drop in performance between the synthetic and real-world evaluation datasets, indicating that the synthetic dataset was not a perfect representation of the real-world data, and that there is still progress to be made in making the synthetic dataset more realistic.
\\

\begin{figure}[htbp]
    \centering
    \includegraphics[width=\linewidth]{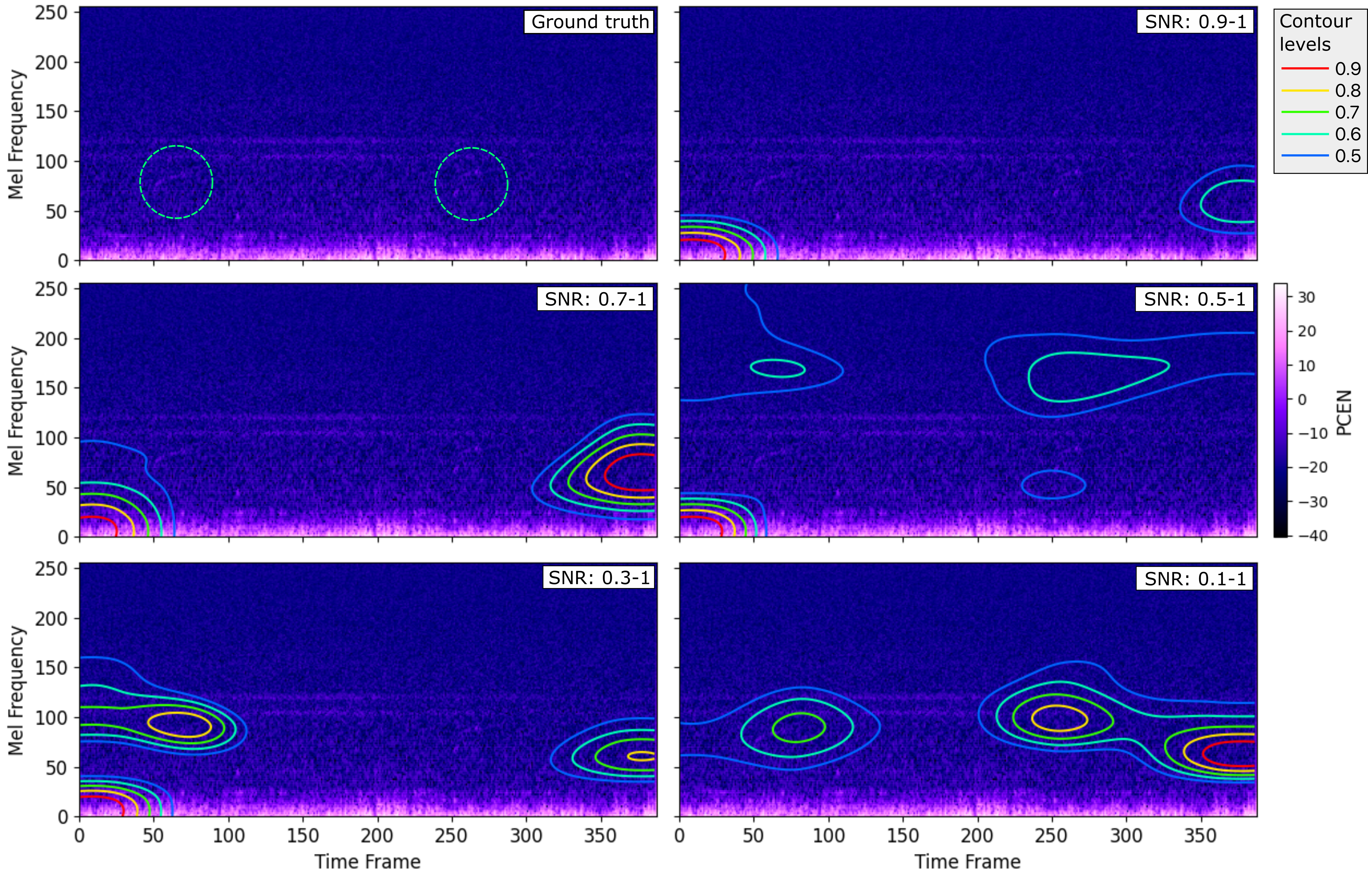}
    \caption{Activation maps of the model's final convolutional layer for a 10-second soundscape segment. The original soundscape (top left) contains two very low SNR \textit{Athene noctua} vocalisations (indicated by green circles). The remaining five images show models trained on synthetic datasets generated with varying SNR ranges.}
    \label{fig:activations}
\end{figure}

We also qualitatively evaluated the effect of varying the SNR range of the synthetic dataset on model performance on low-SNR samples. We evaluated the ability of models trained on synthetic datasets with varying SNR ranges to detect low-SNR samples by comparing the activation maps of the final convolutional layer of the model (Figure \ref{fig:activations}). Only the model trained on the synthetic dataset with a SNR range of -10 to 0 dB (0.1-1 linear scale) was able to positively detect the low-SNR vocalisations with greater than 50\% confidence. Models trained with a higher minimum SNR (e.g. \ref{fig:activations}, top right and mid left) did not show any activation around the vocalisations, indicating that these models were unable to detect the low-SNR vocalisations. Anecdotally, we observed similar results on other low-SNR samples. This suggests that models trained on the synthetic dataset with lower SNR augmentations were more robust to low-SNR samples in real world environment.
\\

\section{Discussion}
Our results demonstrate the effectiveness of the proposed synthetic data generation framework for training robust bioacoustic classification models from very small datasets $(s < 3)$. Model performance on the held-out evaluation data reached an Area Under the Curve of greater than 0.9, indicating that the model successfully generalised to the real-world environment from the completely synthetic dataset. These findings are particularly encouraging given the challenges often faced in bioacoustic machine learning due to limited labelled data \citep{stowellComputationalBioacousticsDeep2022}. Our approach demonstrates that it is possible to create a large, diverse, and functional training dataset from a very small number of isolated vocalisations, potentially overcoming a significant barrier in the field.
\\

Model performance did reduce slightly once the synthetic dataset size exceeded 1000 examples, or if the number of unique isolated vocalisations was reduced below 3. However, the model still achieved high performance even with a single unique isolated vocalisation, strongly suggesting that the model was able to learn generalised features from the synthetic dataset rather than overfitting to augmentation artifacts. Both the reduction in performance after 1000 examples and the performance drop when moving from the synthetic dataset to the real-world evaluation dataset suggest that the synthetic dataset is not a perfect representation of the real-world data. This is likely due to the fact that the synthetic dataset was generated using a limited number of isolated vocalisations, which may not fully capture the diversity of vocalisations present in the real-world environment. The synthetic data generation framework may also introduce biases, and there are likely more subtle domain differences between the synthetic and real-world data that the model is unable to learn. However, synthetic soundscapes are highly realistic and largely free of the perceptual artifacts often associated with generative models \citep{gibbonsGenerativeAIbasedData2024}. Another benefit of our approach compared with generative models is that the synthetic dataset can be generated in a controlled manner, allowing for the downstream model's SNR sensitivity to be tuned. In the same fashion it is also possible to explicitly control the weather or background noise conditions of the synthetic dataset, which may be useful for training models for specific environments.
\\

While our method initially requires a small, one-time manual effort to create an isolated vocalisation dataset, effective models can be trained with as few as one such vocalisation per sonotype. Future automation of isolated vocalisation generation, as explored by \citet{hoffmanSyntheticDataEnables2025}, could further minimize this initial step. A key strength of our approach is its use of vocalisations directly from PAM recordings, thereby avoiding the performance dip often observed when transferring models from focal-recording datasets to the PAM domain \citep{kahlBirdNETDeepLearning2021}. By automating the generation of high-specificity training data, including bounding boxes and masks, our method drastically reduces manual labor. This automation accelerates dataset creation by orders of magnitude and enforces rigorous label consistency, effectively eliminating the observer variability that plagues human annotations. Furthermore, the framework's capacity to synthesize thousands of ecologically plausible training examples from limited source vocalisations leads to robust models with strong generalization potential, which is a significant advantage in bioacoustics where diverse field recordings often hinder model transferability \citep{merchanBioacousticClassificationAntillean2020}. Although our current evaluation uses binary classification for methodological validation, the framework's architecture readily extends to multi-species detection scenarios by synthesizing rich class labels with soundscapes. This positions our approach as particularly valuable for large-scale biodiversity monitoring, addressing the fundamental challenge of simultaneously surveying numerous species across dynamic landscapes in computational ecology.\\

While our results are promising, a few potential limitations and areas for future research should be noted. Firstly, our evaluation approach focuses on binary classification at soundscape-level specificity. While the augmentation method is fully capable of producing richly labelled datasets, future work could explore whether the same trend observed here also holds for higher specificity detection and classification. This would be valuable for species-specific monitoring studies. Additionally, while our augmentation framework aims to create realistic artificial soundscapes, there may be subtle characteristics of real-world recordings that are not fully captured. Further research could investigate ways to incorporate more complex environmental factors into the augmentation process, such as more sophisticated background noise models and species-specific consecutive vocalisation patterns. The accuracy of automatically generated rich labels could also be further developed to potentially improve the consistency of the training target. Finally, while our evaluation dataset represented novel conditions, further testing across a wider range of environments and recording conditions would provide additional insight into the generalisation capabilities of models trained using this approach.\\

In conclusion, our synthetic data augmentation framework addresses three critical bottlenecks in bioacoustic machine learning: (1) the prohibitive cost of manual vocalisation-level annotations, (2) the spectral and temporal homogeneity of conventional training datasets, and (3) fine-grained controllable training dataset generation. By dynamically generating ecologically realistic soundscapes with physically consistent labels, we enable models to learn to detect and classify target vocalisations within complex, previously unseen soundscapes. Crucially, our evaluation demonstrates that performance persists even with very few source isolated vocalisations and with large synthetic datasets, indicating that the data synthesis framework effectively mimics real-world conditions well enough to offset the highly imbalanced and data-limited nature of the PAM domain.\\

As conservation timelines compress and PAM deployments expand into understudied ecosystems, our method offers a critical tool for rapid, cost-effective model adaptation, empowering researchers to extract fine-grained ecological metrics like VAR and possibly density. Future integration of automatic vocalisation isolation, in-context synthesis for novel classes, and more realistic soundscape synthesis will enable fully automated PAM analysis, ultimately advancing computational ecology’s capacity to monitor biodiversity at scale.

\section*{Acknowledgments}
This research was supported by the University of Canterbury Aho Hīnātore Scholarship provided to KS. Field data collection was primarily conducted by Erin Mornin (Department of Biological Sciences, University of Canterbury) as part of a separate project.

\section*{Author Contributions}
KS and SG conceived the ideas and designed the methodology; KS collected the source audio data and implemented the software; KS analysed the data; KS, SG, and TS interpreted the results; KS led the writing of the manuscript. All authors contributed critically to the drafts and gave final approval for publication.

\section*{Conflict of Interest}
The authors declare no conflict of interest.

\section*{Data Availability}
Code developed for data augmentation, preprocessing, model training, and evaluation, along with configuration files and links to source datasets (where publicly available, e.g., Xeno-Canto IDs), are publicly archived at: \url{https://github.com/KasparSoltero/bioacoustic-data-augmentation-small}.

\printbibliography

\end{document}